\documentclass[epj]{svjour}

\input{epsf}
\epsfclipon

\begin{document}

\title{Ground state and glass transition of the RNA secondary structure}

\author{Sheng Hui and Lei-Han Tang
\thanks{\email{lhtang@hkbu.edu.hk}}}
\institute{Department of Physics, Hong Kong Baptist University, Kowloon 
Tong, Hong Kong SAR, China}

\date{\today}

\abstract{
RNA molecules form a sequence-specific self-pairing pattern at low
temperatures. We analyze this problem using a random pairing
energy model as well as a random sequence model that includes
a base stacking energy in favor of helix propagation.
The free energy cost for separating a chain into two equal halves
offers a quantitative measure of sequence specific pairing.
In the low temperature glass phase, this quantity 
grows quadratically with the logarithm of the chain length, but
it switches to a linear behavior of entropic origin in the high
temperature molten phase. Transition between the two phases is
continuous, with characteristics that resemble those of a disordered 
elastic manifold in two dimensions. For designed sequences, however,
a power-law distribution of pairing energies on a coarse-grained level
may be more appropriate. Extreme value statistics arguments then
predict a power-law growth of the free energy cost to break a chain,
in agreement with numerical simulations. Interestingly, the distribution
of pairing distances in the ground state secondary structure follows 
a remarkable power-law with an exponent $-4/3$, independent
of the specific assumptions for the base pairing energies.
}

\maketitle

\section{Introduction}

The three-dimensional structure (i.e. conformation) of 
biomolecules is a fascinating topic due to its fundamental importance
in modern biology\cite{ref_ba}. Link between the structure of a
biopolymer and its sequence information, however, remains at an empirical  
level due to the hitherto unyielding computational 
complexity in predicting the shape of a heterogeneous 
polymer\cite{ref_kad,ref_jno,ref_eis,Baker05,Snow}. At the
heart of the problem is the lack of a general understanding
on the energetics of a collapsed polymer in the presence of
sequence-specific contact energies. Such a situation has
been compared with the low temperature behavior of the spin
glass model\cite{ref_tg2,ref_th}, although the chain constraint and the
unknown nature of sequence specificity may invalidate the
analogy.

In the present paper, we focus on the secondary structure of 
RNA molecules\cite{ref_it,ref_rws,ref_gmb,ref_mlma,ref_rb}. RNA, like DNA, 
is a long chain molecule made of four
different types of nucleotides: adenine (A), uracil (U), guanine (G) 
and cytosine (C). Under normal physiological conditions, 
an RNA molecule folds into a relatively compact shape
which can be loosely described as a mixture of 
double-stranded helical segments (known as stems) and occasional 
single-stranded bulges and hairpins with tertiary contacts.
The helical segments are stabilized by base-pairing and base stacking
which represent dominant contributions to the energy of a folded structure.
Unlike a ds-DNA molecule, however, each helical segment is made of two 
complementary segments from different parts of the same chain, 
running in opposite directions. The matching of bases to form the
Watson-Crick A-U and G-C pairs, and the energetically less favorable
wobble G-U pairs defines the secondary structure of an RNA molecule. 
The problem of RNA secondary
structure prediction is then to find the map of optimal
pairings for a given sequence of the nucleotides (the
primary structure)\cite{ref10}. At finite temperatures, one has to
consider structures that are not necessarily optimal in
energy, but are nevertheless important due to their
configurational entropy.

Compared to protein folding, RNA secondary structure prediction is a 
simpler problem due to the saturation of base-pairing\cite{ref_it}.
In particular, for RNA molecules without the so called ``pseudoknots'',
pairing of bases in an RNA molecule may be
represented by one-dimensional, non-intersecting rainbow
diagrams\cite{ref7}. Thanks to this topological constraint,
the partition function of a chain of $N$ bases
can be determined through an exact dynamic programming
algorithm whose computational complexity scales as
$N^3$\cite{DynProg,ref_rd}. 
Consequently, chains of length up to a few
thousand bases can be readily investigated numerically.

From a statistical mechanics point of view, the key issues 
with regard to RNA secondary structures include
a classification of possible phases of the chain in the
limit $N\rightarrow\infty$, and the characteristics of the
equilibrium structures in each of these phases\cite{ref_rb,ref7}.
At sufficiently high temperatures, it is generally agreed that
the system is in a ``molten phase'' with non-specific base-pairing.
Various statistical properties of this phase, including the
distribution of pairing distances, are known through the analytic
solution to the homopolymer version of the RNA problem\cite{ref12}.
A defining property of the molten phase is the universal amplitude of
the logarithmic excess entropy of a finite chain\cite{ref7}.
As the temperature decreases, a new type of behavior, with properties
typical of disordered systems, is seen. However, many details
remain controversial\cite{ref4,ref5,ref6,Krzakala,MPR}.

Several simplifying models have been introduced in the study of the
low temperature glass phase of an RNA molecule.    
Higgs considered a random heteropolymer model of RNA
secondary structure formation\cite{ref4}. In his model, only 
Watson-Crick pairing is allowed and each such pair is assigned a
negative energy. Through numerical simulations of random
sequences, he observed that the ground state is highly
degenerate and the system at low temperatures exhibits a
broad distribution of the overlap function.
The same conclusion was reached in a
recent work by Pagnani {\it et al.} who also studied the 
molten-to-glass transition\cite{ref5}. The existence of a spin-glass type
ground state in such a model is however disputed by Hartmann.\cite{ref6}
    
Bundschuh and Hwa have recently carried out extensive analytic and
numerical studies of the RNA secondary structure problem\cite{ref7}.
They have discussed in particular the nature and energetics of
low-energy excitations in the glass phase, and presented a proof
for the existence of a finite-temperature glass transition.
They have shown that the scaling of pairing distances in the
glass phase follows a different power from that of the molten phase
(see discussion in Sec. 3.2). In addition, the finite-size correction
to the free energy (termed pinching energy by the authors)
grows as a power-law of the chain length, 
but the exponent is small and nonuniversal.

Krzakala {\it et al.}\cite{Krzakala} introduced an alternative measure
of the sequence-specific pairing which is a characteristic of the
low temperature glass phase. Their approach is based on an analogy
to the directed polymer problem\cite{mezard90} and the replica method. 
Their conclusion on the existence of a finite-temperature glass transition
is in agreement with previous work.

While the quantities introduced by Bundschuh and Hwa, and by 
Krzakala {\it et al.} provide effective measures of the glassy order
in the low temperature phase, there is yet no microscopic understanding 
of the origin of the scale-dependent energies as seen in numerical
work. In particular, there is no compelling reason why power-law
forms are the preferred choice for the observed scale dependence. 
This question is important not only from a theoretical point of view, but also
when considering the effect of sequence mutation and environmental
perturbations (such as tertiary contacts, pseudoknots, and magnesium
ions, etc.) on the RNA secondary structure. Therefore a better 
characterization of the properties of the low-temperature phase is desirable.

The paper is organized as follows. In Sec. 2 we introduce the random
pairing energy model studied in the present work and briefly review the 
numerical scheme used for exact computation of ground state and finite 
temperature properties. Section 3 contains results and analysis
of various properties in the ground state. The behavior of the system
at finite temperatures is discussed in Sec. 4. In Sec. 5 we consider
other specification of the random pairing energy and their effect on the 
properties of the ground state. Section 6 presents a summary and our main 
conclusions.
    
\section{The model and dynamic programming}

The statistical mechanics of the secondary structure of random RNAs
is reviewed in Ref.\cite{ref7}. 
An RNA molecule is defined by its nucleotide sequence.
A secondary structure of the molecule is a pairing pattern of bases 
on the sequence,
where each base (indexed by its position $i$ in the sequence)
has at most one partner. As in most previous studies,
we consider here only secondary structures that obey the ``noncrossing''
constraint, i.e., if base $i$ pairs with base $j>i$, and another base
$k>i$ pairs with base $l>k$, then either 
$i<j<k<l$ (separated) or $i<k<l<j$ (nesting).
This class of structures, which are the most common in nature,
form the configuration space of the RNA secondary structures considered
below.

Realistic prediction of the thermodynamically favored RNA secondary
structures requires a large parameter set derived empirically from 
pains-taking thermodynamic measurements over the years\cite{santalucia}.
Its main purpose is to differentiate accurately
local pairing alternatives. This complication, we believe, is not
necessary for a statistical characterization of the scaling properties
in the low temperature phase and around the glass transition in the random 
sequence ensemble. Instead, we consider here a much simpler model
where the energy of a secondary structure $S$ is given by,
\begin{equation}
E[S]=\sum_{(i,j)\in S}\epsilon_{i,j},
\label{E-model}
\end{equation}
where $\epsilon_{i,j}$ is the pairing energy of base $i$ with base $j$.
The sum is over all base pairings $(i,j)$ of $S$.

To complete the description of the model, we need to assign values to
the pairing energies $\epsilon_{i,j}$ for a given nucleotide sequence.
The standard choice is to make $\epsilon_{i,j}$ dependent on the two
nucleotides involved. 
For the random sequence ensemble, an alternative approach is to
choose $\epsilon_{i,j}$ as independent random variables,
as suggested in Ref.\cite{ref7}. This was motivated at first by
analytical considerations and supported by numerical evidence. 
In fact, the two approaches become quite identical when the alphabet 
size exceeds sequence length, as then every possible pair has a different 
combination of partners for a typical random sequence. 
Considering that, for real RNA, each helical segment typically contains
a consecutive stack of five or more paired bases (with more than
$4^5=1024$ possible sequences on each side), one may view
the second approach as defining a coarse-grained model on the scale
of a helical segment. Previous work on sequence alignment has shown that
the matching energy of two randomly selected sequences follows a
distribution with an exponential tail\cite{Dembo,tang01,YuHwa,hwa03}. 
Thus, as a coarse-grained 
model of RNA secondary structures in the sense described above,
we choose $\epsilon_{i,j}<0$ to be independent random variables
satisfying the distribution,
\begin{equation}
P(\epsilon)=\epsilon_0^{-1}\exp(\epsilon/\epsilon_0),
\label{P-epsilon}
\end{equation}
where $\epsilon_0=1$ sets the only energy scale of the problem.

Due to the noncrossing constraint on the pairing patterns,
the partition function 
\begin{equation}
Z(N)=\sum_{S}\exp(-E[S]/T)
\label{partition_fn}
\end{equation}
of an RNA molecule of $N$ bases at temperature $T$ can be calculated
using a dynamic programming algorithm\cite{DynProg,ref_rd}. 
This is done based on the recursive relation
\begin{equation}
Z_{i,j}=Z_{i,j-1}+\sum_{k=i}^{j-1}Z_{i,k-1}e^{-\epsilon_{k,j}/T}Z_{k+1,j-1}.
\label{Z-iter}
\end{equation}
Here $Z_{i,j}$ denotes the partition function of a contiguous segment 
of the molecule from position $i$ to position $j$.
Starting from the shortest segments of one base each with
$Z_{i,i}=1, i=1,2,\ldots,N$, one obtains the partition function 
$Z(N)\equiv Z_{1,N}$ in $O(N^3)$ elementary computations.
At $T=0$, the following equation can be used instead to calculate the
ground state energies,
\begin{equation}
E_{i,j}={\rm min}_{i\leq k\leq j}\{E_{i,k-1}+E_{k+1,j-1}+\epsilon_{k,j}\},
\label{ground-state}
\end{equation}
where as a convention we set $\epsilon_{i,i}=0$ for all $i$, 
and $E_{i,j}=0$ for $i\geq j$. 

\section{The ground state}

In this section we present numerical results regarding the ground
state of an RNA molecule in the random sequence ensemble.
Chains up to $N=2048$ bases are investigated, with a minimum
of 1000 realizations of the pairing energies. Results for shorter 
chains are obtained as a byproduct in the computation.

\subsection{Ground state energy}

It has been suggested\cite{ref7,Krzakala} 
that the statistical mechanics
of the RNA problem may be closely related to that of a directed
polymer in a disordered medium, which has been studied extensively
in the past\cite{zhang95}. In the latter case, the ground state energy 
of the polymer (or its free energy at finite $T$) 
contains a finite-size correction which grows as a
power of the chain length\cite{krug90,tfw}. This energy is of the
same order as the disorder-induced energy fluctuations, with
an exponent that takes a universal value throughout the low temperature 
phase. It is thus interesting
to examine such corrections for the RNA problem as well.

The origin of an excess energy associated with a chain of finite length
can be appreciated with the help of Fig. 1. Dashed lines in the figure 
indicate pairing of the bases. Cutting the chain in the middle
yields two shorter chains half of the original size.
All pairing patterns of the two shorter chains can be realized 
on the longer chain, but the reverse is not true. 
Therefore the free energy of the chain increases when
it is broken into smaller parts. This property translates directly
to an excess free energy for a chain of finite length.

\begin{figure}
\epsfxsize=\linewidth
\epsfbox{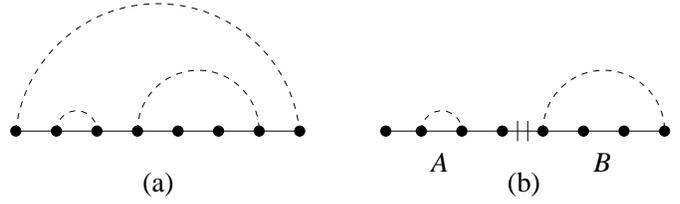}
\caption{Rainbow diagrams illustrating allowed base pairing (dashed line)
on (a) a single chain, and (b) two separate parts when the chain is broken
in the middle. 
}
\label{rainbow}
\end{figure}

The importance of quantifying this excess energy has been stressed
by Bundschuh and Hwa\cite{ref7}. Due to the noncrossing constraint, when
two bases $i$ and $j$ on the chain form a pair, those within the segment 
delimited by the two are only allowed to pair among themselves. Therefore 
any pairing of the bases effectively defines a finite system isolated
from the rest of the chain. For the pairing to be energetically favorable,
its energy $\epsilon_{i,j}$ must offset the energy cost for splicing out the 
segment inbetween. Arguments along this line can be used to discuss
the stability of a given state as done in Ref. \cite{ref7} to
construct a lower bound on the glass transition temperature.

Here we examine not only the average value but also the distribution
of the excess energy as a function of the chain length.
Due to the statistical fluctuations in the bond energies, the total
ground state energy $E(N)$ of a chain of length $N$ has a fluctuation
proportional to $N^{1/2}$. This background fluctuation can be eliminated
using the construction shown in Fig. 1(b). A chain of length $2N$
is formed by joining two chains $A$ and $B$, each of length $N$. 
Let $\Delta E_N\equiv E_{1,N}+E_{N+1,2N}-E_{1,2N}$ be the energy gained
when bases on chain $A$ are allowed to pair with bases on chain $B$
to form the ground state of the full chain.
Apart from the energy of a single pair,
this quantity is identical to the pinching energy introduced
in Ref.\cite{ref7}. Chemically, it can be considered as the 
heat of ``reaction'' that brings the two halves together.
Obviously, $\Delta E_N$ is typically
positive but may happen to be zero when the ground state of the full
chain breaks into two independent halves.

Figure 2(a) shows the normalized distribution $P(\Delta E,N)$ of 
$\Delta E_N$ for $N=2,4,8,\ldots, 1024$ on semi-log scale. 
As $N$ increases, the peak of each curve shifts to the right while 
at the same time its width also increases. In addition, there is a 
finite statistical weight $P_0(N)\sim N^{-4/3}$ at $\Delta E=0$ 
[see Fig. 4(a)]
Figure 2(b) shows a scaling plot of the distributions.
Here $\langle\Delta E_N\rangle$ and 
$W_N=\sqrt{\langle\Delta E_N^2\rangle-\langle\Delta E_N\rangle^2}$ 
denote the mean value and standard deviation of $\Delta E_N$, respectively.
Convergence to a limiting form at $N=\infty$ starts from the
middle of each curve and gradually extends over to the wings. 
Interestingly, the tail of the distributions at large $\Delta E$
decays as a simple exponential. 

\begin{figure}
\epsfxsize=\linewidth
\epsfbox{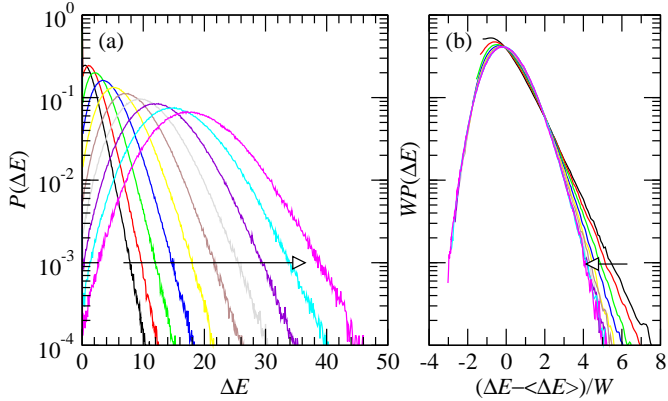}
\caption{(a) Distribution of the excess energy $\Delta E_N$ of a finite
chain for $N=2,4,\ldots,1024$. (b) Convergence to a limiting form with
zero mean and unit variance. Arrows indicate the direction of increasing
$N$.
}
\label{DeltaE_dis}
\end{figure}

Figure 3 shows $\langle\Delta E_N\rangle$ and $W_N$ against
$\ln N$. It is evident that the two quantities are not proportional
to each other, i.e., the distributions shown in Fig. 2 can not be
collapsed with a single energy scale at each $N$. Nevertheless,
the data can be represented nicely by a quadratic function in $\ln N$
for $\langle\Delta E_N\rangle$, and a linear function in $\ln N$
for $W$. This suggests that the disorder averaged ground state energy can
be written in the form,
\begin{equation}
\langle E(N)\rangle=e_0N+a+b\ln N+c\ln^2 N,
\label{finite-size}
\end{equation}
where $e_0$ is the energy per base in the infinite size limit.
From the fit we obtain $a=0.81, b=1.28,$ and $c=0.26$.

\begin{figure}
\epsfxsize=\linewidth
\epsfbox{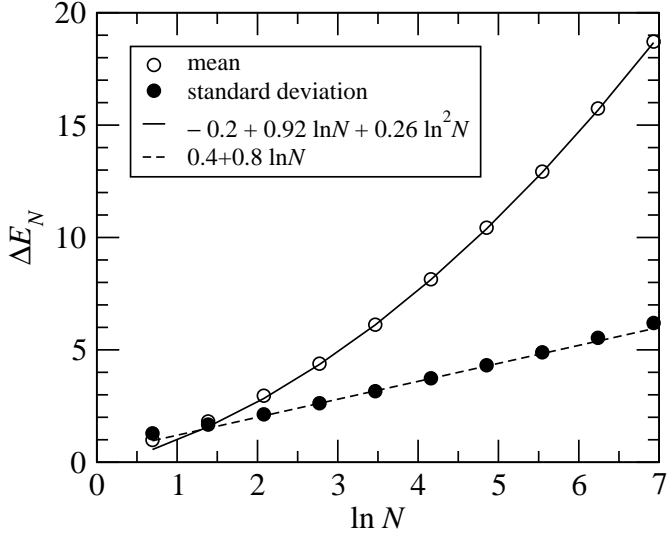}
\caption{Mean value and standard deviation of $\Delta E_N$ against $\ln N$.
Solid and dashed lines represent polynomial fits to the data.
}
\label{gs_energy}
\end{figure}

Although the logarithmic form (\ref{finite-size}) fits the data
nearly perfectly, a power-law dependence can not be ruled out
based on the numerical data alone.\cite{ref7,Krzakala} 
Previously, Bundschuh and Hwa\cite{ref7} made the suggestion that 
the logarithmic size dependence is more appealing given the
small and nonuniversal exponent obtained from various models.
Here we show that the difference in behavior
for $\langle\Delta E_N\rangle$ and $W_N$ is still consistent with a
single energy scale at each $N$ which grows as $\ln N$. 
In such a scenario, the $\ln^2 N$ term
arises naturally as $\langle\Delta E_N\rangle$ also contains 
contributions from smaller scales. Specifically,
with reference to Fig. 1(b), we may first group bases on either side 
of the break into zones that are evenly spaced on a logarithmic 
scale, according to their distance $R$ (measured in terms of
number of bases along the chain) to the breaking point.
Let $l=\ln R$ be the index of zone $l$ with a width of order $R$,
and suppose that the typical
total energy increase of bases in the zone caused by the break is
proportional to $\ln R$. Adding up contributions up to $l=\ln N$
yields the desired $\ln^2N$ dependence. In comparison, fluctuations
of $\Delta E_N$ do not contain this cumulative effect.

The $\ln N$ energy scale may be motivated from the extreme value 
statistics argument developed in Refs.\cite{ref7,Dembo,tang01,YuHwa,hwa03}.
According to Eq. (\ref{ground-state}), when a new base is added
to the end of a chain of length $N$, optimal pairing with interior bases
is determined by the competition between the energy cost for
perturbing the existing ground state [i.e., the first two terms
on the right-hand-side of Eq. (\ref{ground-state})], and the
energy gain from the newly formed pair. The perturbation is at its weakest
when the partner base $k$ is located at either end of the chain. 
In such a situation, the number of bases at such distances is small, 
so the available choices for the bond energy $\epsilon_{k,j}$ are
rather limited. On the other hand,
when $k$ resides in the middle of the chain, the perturbation is the
strongest, but then there are of order $N$ choices for $\epsilon_{k,j}$
which, according to the extreme value statistics, yield a typical energy
gain proportional to $\ln N$. If $\Delta E_N$ grows as a power of $N$,
this energy gain will not be enough to offset the energy cost associated
with the perturbation. Consequently pairing with a base in the middle
of the chain is extremely unlikely. This however would imply that
the chain contains almost exclusively short-distance pairs. 
If this were the case, the system would have a finite correlation length
and a bounded $\Delta E_N$, which contradicts our original assumption.
Self-consistency thus requires $\Delta E_N$ to grow slower than any power
of $N$, but at least as fast as $\ln N$. 

\subsection{Pairing statistics}

In addition to the ground state energy, we have examined the statistics
of pairing distance $d$ in the ground state. When two bases $i$
and $j>i$ form a pair, their pairing distance is defined as $d_{ij}=j-i$.
In fact, for a chain of length $N$, pairs of size $d$ are equivalent
to pairs of size $N-d$. This becomes evident if we join the two ends
of the chain to form a circle, in which case distance between the two bases
is given by the smaller of $d$ and $N-d$. Let $P_{s}(d)$ be the distribution 
of $d$. Symmetry then yields $P_s(d)=P_s(N-d)$. 
Figure 4(a) shows the distribution $P_s(d)$ for a chain of length 
$N=2048$, averaged over 1000 realizations of the $\epsilon_{i,j}$'s.
From the plot, we see that, apart from those data points near
$d\simeq N/2$ which are influenced by finite-size effects,
$P_s(d)$ follows a power-law function $d^\eta$ with an exponent $\eta=-4/3$.
Also shown in the figure is $P_0(N)$, the probability that
the ground state breaks into two independent halves as shown in
Fig. 1(b), against $N$ which has a similar behavior.
Note that, for each $N$, $P_0(N)$ is the same as $P(\Delta E)$
at $\Delta E=0$ [cf. Fig. 2(a)].

\begin{figure}
\epsfxsize=\linewidth
\epsfbox{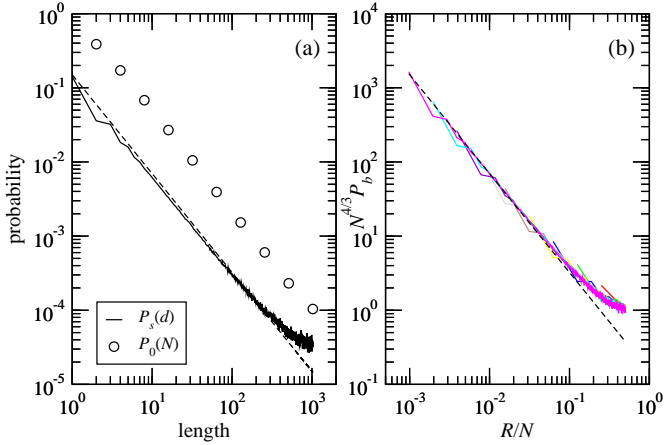}
\caption{Pairing statistics in the ground state. (a) Distribution 
$P_s(d)$ of pairing distance $d$ at $N=2048$ (solid line). 
Also shown is $P_0(N)$ against $N$ (circles).
(b) A scaling plot of the distribution of
the optimal break point for $N=2, 4,\ldots, 1024$. Dashed line
in both figures indicates a power-law function with an exponent
$\eta=-4/3$.
}
\label{split}
\end{figure}

We have also examined the statistics of the location of base $k$
where the minimum on the right hand side of Eq. (\ref{ground-state})
is achieved. Let $R$ be the distance of this base to its partner
base $j$. The distribution $P_b(R,N)$, with $N$ being the 
length of the interval $(i,j)$, is shown in Fig. 4(b) using the
scaled variables. Here $N=2,4,8,\ldots, 1024$. From the data collapse
we conclude that $P_b(R,N)$ obeys scaling,
\begin{equation}
P_b(R,N)=N^{-4/3}\Phi(R/N),
\label{P_b_scaling}
\end{equation}
where $\Phi(x)\sim x^{-4/3}$ for $x\ll 1$.

The scaling properties of base pairing in the 
ground state as discussed above are consistent with the roughness
of ``mountain diagrams'' introduced in Ref.\cite{ref7}.
In the latter representation, a given secondary structure
is mapped to a height profile following a simple rule:
starting from one end of the chain, say $i=0$ with
$h_0=0$, one proceeds successively to the right, setting
$h_i=h_{i-1}+1$ ($h_i=h_{i-1}-1$) if 
base $i$ is paired with base $j>i$ ($j<i$), and $h_i=h_{i-1}$
if base $i$ is unpaired.
Bundschuh and Hwa have shown that the average value of
$h_i$ as defined above grows as a power-law of the chain length
$N$, $\overline{h}\sim N^\zeta$, where the ``roughness exponent''
$\zeta=\zeta_g= 0.67 \pm 0.02$, considerably larger than
its value $\zeta_0=1/2$ in the molten phase. As shown in
Ref.\cite{Krzakala}, the two exponents $\zeta$ and $\eta$ satisfy
a general scaling relation,
\begin{equation}
\zeta=2+\eta.
\label{scaling-relation}
\end{equation}
Equation (\ref{scaling-relation}) holds both in the ground state and
in the molten phase, where $\eta_0=-3/2$ has been calculated exactly.

\section{Finite temperature properties and the glass transition}

At finite temperatures, one needs to consider the entropy associated
with alternative pairing to determine the equilibrium structure
of an RNA molecule. Comparing Eq. (\ref{Z-iter}) with 
Eq. (\ref{ground-state}), we see that qualitatively two types of 
behavior can be distinguished: (i) only one or a few terms on the
right hand side of (\ref{Z-iter}) contribute to $Z_{i,j}$,
in which case the situation is similar to that of the ground state;
(ii) the number of terms that contribute significantly to $Z_{i,j}$ grows 
with the chain length, in which case pairing becomes non-specific and
one is in the molten phase. 

As a quantitative criterion that differentiates the two situations,
Bundschuh and Hwa\cite{ref7} proposed to examine the size dependence of 
the free energy cost for imposing a pairing (termed ``pinching'').
At sufficiently high temperatures, the pinching free energy
$\Delta F$ grows with the pair size $N$ as ${3\over 2}T\ln N$, 
and hence is purely entropic. Based on an estimate of the energy
gain for the best matched pair forbidden by the pinch,
Bundschuh and Hwa argued that this behavior cannot continue
below a certain temperature, and hence a glass transition is expected
to take place. Therefore the size-dependence of $\Delta F$ can be
used to locate the phase transition point.

Following this line of thinking, we consider the statistics
of $\Delta F_N\equiv T\ln (Z_{1,2N}/Z_{1,N}Z_{N+1,2N})$ which
is the finite temperature analog of $\Delta E_N$ defined in the
previous section. Figure 5(a) shows the mean value of $\Delta F_N$
against $\ln N$, with the high temperature behavior\cite{ref7}
$(3/2)T\ln N$ subtracted from the data. For $T=1.25$ and below,
there is a clear upward curvature in each data set, indicating
presence of a $\ln^2 N$ term, though its amplitude decreases with
increasing temperature.  At $T=1.5$ and $1.75$, however,
deviations from the expected high temperature behavior is weak.
Figure 5(b) shows the standard deviation 
$W_{F,N}\equiv\sqrt{\langle \Delta F_N^2\rangle-\langle\Delta F_N\rangle^2}$
against $\ln N$. Using data points at large $N$, we extracted
the slope $A(T)$ of each curve and plotted the result against $T$ as in
the inset. The result can be summarized as,
\begin{equation}
W_{F,N}=
\cases{
A(T)\ln N+B(T),&$T<T_g$;\cr
B(T),&$T>T_g$.\cr}
\label{width-deltaF}
\end{equation}
Here $A(T)=A_0(T-T_g)^2$, with $T_g\simeq 1.7$.

\begin{figure}
\epsfxsize=\linewidth
\epsfbox{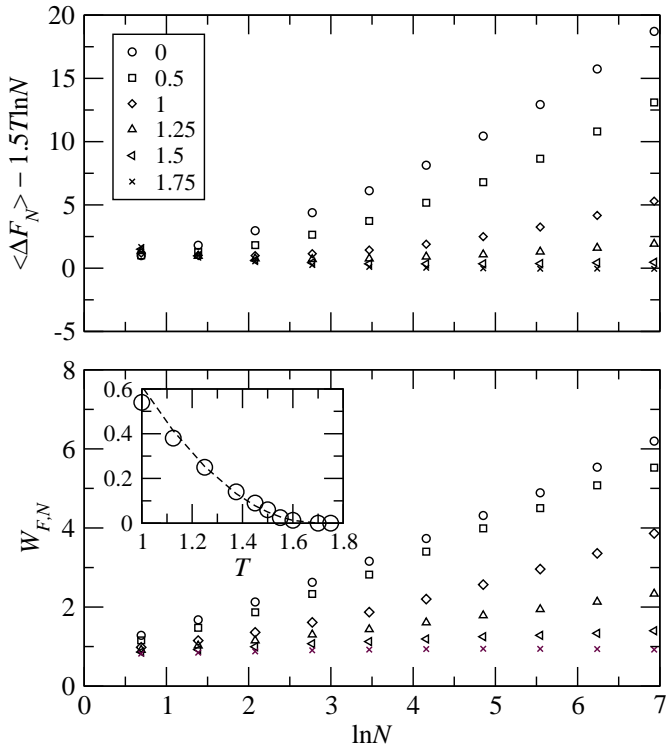}
\caption{(a) Mean and (b) standard deviation of $\Delta F_N$ against 
$\ln N$ at a set of temperatures below and above the glass transition.
Inset in (b) shows the slope of each curve (and additional ones not shown) 
against temperature. Dashed line is a quadratic fit with
$T_g=1.7$.
}
\label{coeff}
\end{figure}

The simple functional forms which fit well the numerical data strongly
suggest that there is an underlying simple mathematical structure.
It is quite conceivable that a renormalization group theory, similar
to the one introduced in Ref. \cite{tang01} for the unbinding transition
of two heteropolymers, can be devised. In the absence of such
a theory, Eq. (\ref{width-deltaF}) should merely be considered as a
convenient representation of numerical data.

\section{Other models for the pairing energy}

We have argued in Sec. 2 that Eq. (\ref{P-epsilon}) provides
a generic description of the distribution of pairing energies
on a coarse-grained scale. To verify this hypothesis, and to
find out to what extent the scaling properties obtained under 
(\ref{P-epsilon}) remain universal, we consider in this section
other forms of the pairing energy, and carry out a comparative
study of their ground state properties.

\subsection{A sequence-based model}

To get a flavor of the similarities and differences between
random sequence models (with $N$ random variables) 
and random pairing energy models (with $N^2$ random variables),
we consider here a simplified four-nucleotide model incorporating
the essential features of base-pairing energetics\cite{ref10}.
In addtion to the Watson-Crick A-U and G-C pairs, we allow the less
favorable G-U pair. A stacking energy is included for
the propagation of short helices, i.e., if two consecutive bases
$i$ and $i+1$ pair with $j$ and $j-1$, respectively, an additional
energy $E_s$ is gained. The minimal hairpin loop length is set
at 4 nt. Results presented below are for the following
choice of energy parameters:
$E_s=-3; E_{\rm GC}=-3-E_s; E_{\rm AU}=-2-E_s; E_{\rm GU}=-1-E_s$. 
With this specification, isolated
pairings are disfavored. In the ground state, the typical length of a 
helix is about five base pairs for a random sequence.

Figure 6 shows the mean value and standard deviation of $\Delta E_N$
against $\ln N$ for the random sequence ensemble. The behavior is very similar
to that of Fig. 3. We have also examined the statistics of the pairing
distances whose distribution fits well to the scaling form (\ref{P_b_scaling})
with $\eta=-4/3$. These and other properties of the ground state will
be reported in detail elsewhere.

\begin{figure}
\epsfxsize=\linewidth
\epsfbox{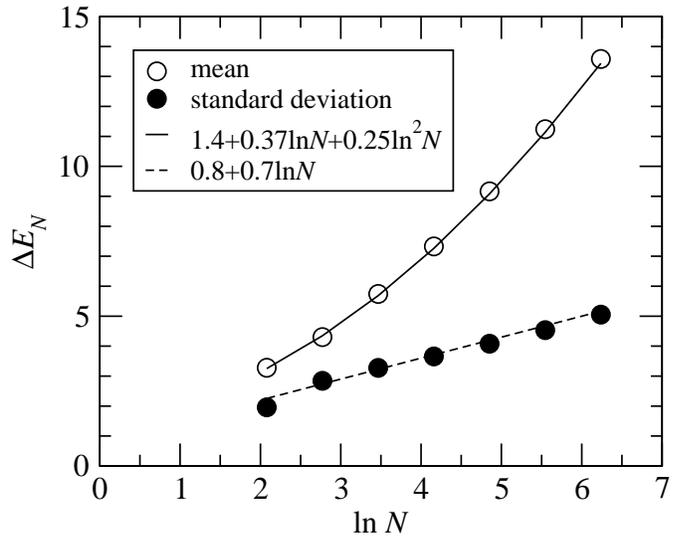}
\caption{Mean value and standard deviation of the excess energy of
a finite chain in the random sequence ensemble.
}
\label{fig6}
\end{figure}

\subsection{Power-law distribution of the pairing energy}

As we mentioned above, 
the logarithmic size dependence of pairing energies is a generic
feature of matching statistics for random sequences. 
Through evolution,
however, sequences that lead to more stable structures may be selected
for functional advantages, including possibly RNA's with longer
matched segments. Indeed, the secondary structure of many real
RNA's show extended stretches of duplices which are not expected
of a random sequence. This observation motivates us to examine
the ground state energetics and pairing pattern under a power-law
distribution of the pairing energies,
\begin{equation}
P(\epsilon)=\alpha|\epsilon|^{-\alpha-1},\qquad\qquad\epsilon\leq -1.
\label{power-law}
\end{equation}

Figure 7 shows the mean and standard deviation of $\Delta E_N$ against
$N$ for $\alpha = 2, 3$, and 4. At sufficiently large $N$, the two 
quantities become proportional to each other, indicating a single energy 
scale $\Delta E_N\sim N^\omega$. The exponent $\omega$ can be related
to $\alpha$ from the following consideration.
On a chain of length $N$, there are $N(N-1)/2$ possible pairings.
The lowest pairing energy $\epsilon_{\rm min}$ is determined by 
the condition $N^2|\epsilon_{\rm min}|^{-\alpha}\sim 1$. Hence
$\epsilon_{\rm min}\sim -N^{2/\alpha}$. Assuming the energy cost for
breaking a chain into two halves is dominated by $\epsilon_{\rm min}$ 
of the strongest bond, we obtain,
\begin{equation}
\omega=2/\alpha,
\label{omega}
\end{equation}
which agrees well with the numerical data.

\begin{figure}
\epsfxsize=\linewidth
\epsfbox{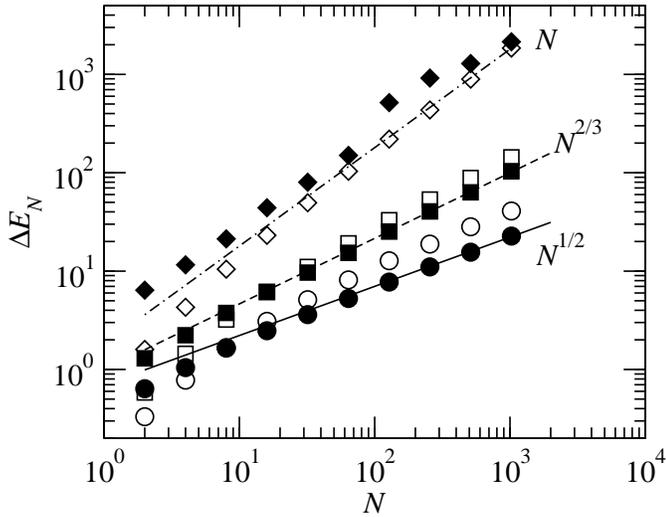}
\caption{Power-law distribution of the pairing energies:
$\alpha=2$ (diamond), $\alpha=3$ (square), and $\alpha=4$ (circle).
Open symbols represent the mean value of $\Delta E_N$ while filled symbols
represent its standard deviation. Solid, dashed, and dot-dashed lines
indicate corresponding power-laws.
}
\label{fig7}
\end{figure}

We have also investigated the distribution of the pairing distance under
Eq. (\ref{power-law}). Interestingly, the results are quite insensitive
to the value of $\alpha$, and the exponent $\eta$ is unchanged from
its value $-4/3$ under (\ref{P-epsilon}). Figure 8 shows a representative
case at $\alpha=2$. The result is nearly identical to Fig. 4(b).
The good data collapse confirms validity of
Eq. (\ref{P_b_scaling}) for power-law pairing energies.

\begin{figure}
\epsfxsize=\linewidth
\epsfbox{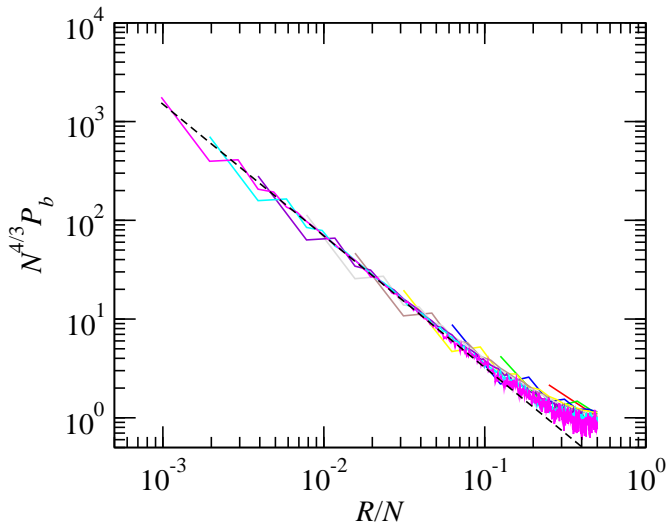}
\caption{Scaling plot of the distribution of the optimal break point
along a chain of length $N$ for the power-law pairing energy model.
}
\label{fig8}
\end{figure}

\section{Summary and conclusions}

In this paper we have investigated properties of the ground state
of RNA secondary structure models, and the transition to the
molten phase at a finite temperature. We have focused our attention
on the excess energy $\Delta E_N$ (and the similarly defined excess
free energy $\Delta F_N$) of a chain of $N$ nucleotides due to the
presence of boundaries. Since any pairing of two bases automatically
splices out a finite segment of the chain, this excess energy
defines the characteristic energy scale for the
competition between base pairing on a given length scale (measured
by the number of nucleotides inbetween) and the adjustments in the
secondary structure necessary in order to accommodate the pairing.
From numerical investigations of a random energy model with
exponential distribution of pairing energies, and a random sequence
model with more realistic base pairing and base stacking energies,
we have established that $\Delta F_N$ has a fluctuation
proportional to $\ln N$ in the entire low
temperature glass phase. The mean value of $\Delta F_N$, on the
other hand, acquires a $\ln^2N$ term due to accumulation of
contributions from smaller scales. 

As temperature increases towards the transition, we observe that
fluctuations of $\Delta F_N$, or equivalently, variation of
the free energy cost to accommodate an inserted pair [i.e., 
the relative strength of different terms in the sum in 
Eq. (\ref{Z-iter})] decreases, hence base pairing
becomes less specific. As $T\rightarrow T_g$, the amplitude
of the $\ln N$ term vanishes quadratically with the distance
to the critical point. This behavior is in striking resemblance
to the glass transition of an elastic manifold in two
dimensions subject to a random, uncorrelated potential
\cite{Zeng,leDoussal}.
It would be interesting to quantify this connection mathematically.

We have also studied scaling properties in the ground state
under a power-law distribution of the pairing energies $\epsilon_{i,j}$.
Such distributions may be encountered in a coarse-grained description
of real RNA molecules with sequence design\cite{ref_rm,Isaacs}. 
In this case, $\Delta E_N$
competes with the strongest bond on the chain. Based on
extremal statistics arguments, we were able to express
the exponent $\omega$ characterizing the power-law growth of 
$\Delta E_N$ with $N$ in terms of the exponent $\alpha$ for
the power-law distribution of $\epsilon_{i,j}$. This relation is
verified by numerical data.

Geometrical properties of base pairing in the RNA secondary structure
can be characterized with the distribution of pairing distances.
Our studies of the ground state shows that this distribution is
well described by a power-law decreasing function with an 
exponent $\eta=-4/3$, in agreement with previous findings\cite{ref7}. 
This behavior is surprisingly insensitive to
the models used for the bond energies. In the molten phase, however,
it takes the value $\eta_0=-3/2$. It would be desirable to 
find an analytic foundation for these observations.

\begin{acknowledgement}
LHT would like to thank Ralf Bundschuh and Terry Hwa
for many interesting discussions on the problem.
Research is supported in part by the Research Grants Council of the 
Hong Kong SAR under grant HKBU 2061/01P. Computations were carried 
out at HKBU's High Performance Cluster Computing Centre Supported by 
Dell and Intel.
We gratefully acknowledge the support of the BIOSUPPORT project
(http://bioinfo.hku.hk) for providing bioinformatics resources and
computational services from the HKU Computer Centre.

\end{acknowledgement}

{\it Note added:} after submission of the paper we became aware of
a manuscript by M. L\"assig and K. J. Wiese\cite{Lassig} where a 
field-theoretic renormalization group treatment of the glass transition 
is presented. The analysis has been further refined\cite{David} and yielded
a value $\zeta_g\simeq 0.64$ at the glass transition.


\begin{thebibliography}{99}

\bibitem{ref_ba} B. Alberts, A. Johnson, J. Lewis, M. Raff, K. Roberts,
P. Walter, {\it Molecular Biology of the Cell} (Garland Science,
New York, 2002).
\bibitem{ref_kad} K.A. Dill, S. Bromberg, K. Yue, K.M. Fiebig, D.P. Yee, 
P.D. Thomas, H.S. Chan, Protein Sci. {\bf 4}, 561 (1995).
\bibitem{ref_jno} J.N. Onuchic, Z. Luthey-Schulten, P.G. Wolynes, 
Annu. Rev. Phys. Chem. {\bf 48}, 545 (1997).
\bibitem{ref_eis} E.I. Shakhnovich, Curr. Opin. Struct. Biol. {\bf 7}, 
29 (1997).
\bibitem{Baker05} O. Schueler-Furman, C. Wang, P. Bradley, K. Misura, 
D. Baker, Science {\bf 310}, 638 (2005).
\bibitem{Snow} C.D. Snow, E.J. Sorin, Y.M. Rhee, V.S. Pande,
Annu Rev Biophys Biomol Struct. {\bf 34}, 43 (2005).
\bibitem{ref_tg2} For a review, see T. Garel, H. Orland, E.  Pitard,
in {\it Spin  Glasses  and Random Fields}, A.P. Young Ed. 
(World Scientific, 1998), p. 387.
\bibitem{ref_th} T. Hwa, Nature {\bf 399}, 17 (1999).
\bibitem{ref_it} I. Tinoco, Jr., C. Bustamante, 
J. Mol. Biol. {\bf 293}, 271 (1999). 
\bibitem{ref_rws} {\it RNA Structure and Function}, Ed. by R.W. Simons, 
M. Grunberg-Manago (Cold-Spring Harbor 1998).
\bibitem{ref_gmb} {\it Nucleic Acids in Chemistry and Biology}, Ed. by 
G.M. Blackburn, M.J. Gait (IRL Press, Oxford, 1990).
\bibitem{ref_mlma} M.L.M. Anderson, {\it Nucleic Acid Hybridization}
(Springer, New York, 1998).
\bibitem{ref_rb} For a recent review see R. Bundschuh, U. Gerland, 
Eur. Phys. J. E {\bf 19}, 319 (2006).
\bibitem{ref10} M. Zuker,D. Sankoff, Bull. Math. Biol. {\bf 46}, 591
(1984); M. Zuker, Science {\bf 244}, 48 (1989).
\bibitem{ref7} R. Bundschuh, T. Hwa, Phys. Rev. Lett. {\bf 83}, 1479
(1999); Phys. Rev. E {\bf 65}, 031903 (2002).
\bibitem{DynProg} R. Nussinov, G. Pieczenik, J.R. Griggs, D.J. Kleitman, 
SIAM J. Appl. Math. {\bf 35}, 68 (1978).
\bibitem{ref_rd} R. Durbin, S.R. Eddy, A. Krogh, G. Mitchison, 
{\it Biological Sequence Analysis}
(Cambridge University Press, Cambridge, England, 1998).
\bibitem{ref12} P.-G. de Gennes, Biopolymers {\bf 6}, 715 (1968).
\bibitem{ref4} P.G. Higgs, Phys. Rev. Lett. {\bf 76}, 704 (1996);
Q. Rev. Biophys. {\bf 33}, 199 (2000).
\bibitem{ref5} A. Pagnani, G. Parisi, F. Ricci-Tersenghi,  Phys.
Rev. Lett. {\bf 84}, 2026 (2000).
\bibitem{ref6} A.K. Hartmann, Phys. Rev. Lett. {\bf 86}, 1382 (2001). 
\bibitem{Krzakala} F. Krzakala, M. M\`ezard, M. M\"uller, 
Europhys. Lett. {\bf 57}, 752 (2002); 
M. M\"uller, F. Krzakala, M. M\`ezard, 
Eur. Phys. J. E {\bf 9}, 67 (2002); 
M. M\"uller, Phys. Rev. E {\bf 67}, 021914 (2003).
\bibitem{MPR} E. Marinari, A. Pagnani, F. Ricci-Tersenghi, 
Phys. Rev. E {\bf 65}, 041919 (2002).
\bibitem{mezard90} M. M\'ezard, J. Physique {\bf 51}: 1831 (1990).
\bibitem{santalucia} J. SantaLucia Jr., Proc. Nat. Acad. Sci. USA
{\bf 95}, 1460 (1998).
\bibitem{Dembo} S. Karlin, A. Dembo,
Adv.\ Appl.\ Probab. {\bf 24}, 113 (1992).
\bibitem{tang01} L.-H. Tang, H. Chat\'e, Phys. Rev. Lett. {\bf 86}, 830
(2001).
\bibitem{YuHwa} Y.-K. Yu, T. Hwa, J.\ Comp.\ Biol. {\bf 8}, 249 (2001). 
\bibitem{hwa03} T. Hwa, E. Marinari, K. Sneppen, L.-H. Tang,
Proc. Nat. Acad. Sci. {\bf 100}, 4411 (2003).
\bibitem{zhang95} T. Halpin-Healy, Y.-C. Zhang,
Phys. Rep. {\bf 254}, 215 (1995).
\bibitem{krug90} J. Krug, P. Meakin, J. Phys. A
{\bf 23}, L987 (1990).
\bibitem{tfw} L.-H. Tang, B.M. Forrest, D.E. Wolf, Phys. Rev. A
{\bf 45}, 7162 (1992).
\bibitem{Zeng} C. Zeng, P.L. Leath, T. Hwa, Phys. Rev. Lett. {\bf 83},
4860 (1999).
\bibitem{leDoussal} D. Carpentier, P. Le Doussal, Phys. Rev. B {\bf 55},
12128 (1997); and references therein.
\bibitem{ref_rm} R. Mukhopadhyay, E. Emberly, C. Tang, N.S. Wingreen, 
Phys. Rev. E {\bf 68}, 041904 (2003).
\bibitem{Isaacs} F.J. Isaacs, D.J. Dwyer, J.J. Collins,
Nat. Biotechnol. {\bf 24}, 545 (2006).
\bibitem{Lassig} M. L\"assig, K. J. Wiese, Phys. Rev. Lett.
{\bf 96}, 228101 (2006).
\bibitem{David} F. David, K. J. Wiese, q-bio.BM/0607044.

\end{thebibliography}
\end{document}